\input harvmac.tex
\input epsf
\noblackbox

\newcount\figno
 \figno=0
 \def\fig#1#2#3{
\par\begingroup\parindent=0pt\leftskip=1cm\rightskip=1cm\parindent=0pt
 \baselineskip=11pt
 \global\advance\figno by 1
 \midinsert
 \epsfxsize=#3
 \centerline{\epsfbox{#2}}
 \vskip 12pt
 {\bf Fig.\ \the\figno: } #1\par
 \endinsert\endgroup\par
 }
 \def\figlabel#1{\xdef#1{\the\figno}}

\newdimen\tableauside\tableauside=1.0ex
\newdimen\tableaurule\tableaurule=0.4pt
\newdimen\tableaustep
\def\phantomhrule#1{\hbox{\vbox to0pt{\hrule height\tableaurule width#1\vss}}}
\def\phantomvrule#1{\vbox{\hbox to0pt{\vrule width\tableaurule height#1\hss}}}
\def\sqr{\vbox{%
  \phantomhrule\tableaustep
  \hbox{\phantomvrule\tableaustep\kern\tableaustep\phantomvrule\tableaustep}%
  \hbox{\vbox{\phantomhrule\tableauside}\kern-\tableaurule}}}
\def\squares#1{\hbox{\count0=#1\noindent\loop\sqr
  \advance\count0 by-1 \ifnum\count0>0\repeat}}
\def\tableau#1{\vcenter{\offinterlineskip
  \tableaustep=\tableauside\advance\tableaustep by-\tableaurule
  \kern\normallineskip\hbox
    {\kern\normallineskip\vbox
      {\gettableau#1 0 }%
     \kern\normallineskip\kern\tableaurule}%
  \kern\normallineskip\kern\tableaurule}}
\def\gettableau#1 {\ifnum#1=0\let\next=\null\else
  \squares{#1}\let\next=\gettableau\fi\next}

\tableauside=1.0ex \tableaurule=0.4pt


\def\inbar{\,\vrule height1.5ex width.4pt depth0pt}
\def\IC{\relax\hbox{$\inbar\kern-.3em{\rm C}$}}

\writedefs

\lref\Amariti{
  A.~Amariti, L.~Girardello and A.~Mariotti,
  ``On meta-stable SQCD with adjoint matter and gauge mediation,''
  arXiv:hep-th/0701121.
}

\lref\Aharony{
  O.~Aharony and N.~Seiberg,
  ``Naturalized and simplified gauge mediation,''
  JHEP {\bf 0702}, 054 (2007), 
  arXiv:hep-ph/0612308.
}

\lref\Csaki{
  C.~Csaki, Y.~Shirman and J.~Terning,
  ``A simple model of low-scale direct gauge mediation,''
  arXiv:hep-ph/0612241.
}

\lref\Murayamab{
  H.~Murayama and Y.~Nomura,
  ``Simple scheme for gauge mediation,''
  arXiv:hep-ph/0701231.
}

\lref\Murayamaa{
  H.~Murayama and Y.~Nomura,
  ``Gauge mediation simplified,''
  arXiv:hep-ph/0612186.
}

\lref\Bousso{
  R.~Bousso and J.~Polchinski,
  ``Quantization of four-form fluxes and dynamical neutralization of the
  cosmological constant,''
  JHEP {\bf 0006}, 006 (2000), 
  arXiv:hep-th/0004134.
}

\lref\ISS{
  K.~Intriligator, N.~Seiberg and D.~Shih,
  ``Dynamical SUSY breaking in meta-stable vacua,''
  JHEP {\bf 0604}, 021 (2006), 
  arXiv:hep-th/0602239.
}

\lref\Oogurib{
  H.~Ooguri and Y.~Ookouchi,
  ``Meta-stable supersymmetry breaking vacua on intersecting branes,''
  Phys.\ Lett.\  B {\bf 641}, 323 (2006), 
  arXiv:hep-th/0607183.
}

\lref\Ooguria{
  H.~Ooguri and Y.~Ookouchi,
  ``Landscape of supersymmetry breaking vacua in geometrically realized gauge
  theories,''
  Nucl.\ Phys.\  B {\bf 755}, 239 (2006), 
  arXiv:hep-th/0606061.
}

\lref\kachru{
  R.~Argurio, M.~Bertolini, S.~Franco and S.~Kachru,
  ``Metastable vacua and D-branes at the conifold,''
  arXiv:hep-th/0703236.
}

\lref\Franco{
  S.~Franco, I.~Garcia-Etxebarria and A.~M.~Uranga,
  ``Non-supersymmetric meta-stable vacua from brane configurations,''
  JHEP {\bf 0701}, 085 (2007), 
  arXiv:hep-th/0607218.
}

\lref\Bena{
  I.~Bena, E.~Gorbatov, S.~Hellerman, N.~Seiberg and D.~Shih,
  ``A note on (meta)stable brane configurations in MQCD,''
  JHEP {\bf 0611}, 088 (2006), 
  arXiv:hep-th/0608157.
}

\lref\KKLT{
  S.~Kachru, R.~Kallosh, A.~Linde and S.~P.~Trivedi,
  ``De Sitter vacua in string theory,''
  Phys.\ Rev.\  D {\bf 68}, 046005 (2003), 
  arXiv:hep-th/0301240.
}

\lref\KOO{
 R.~Kitano, H.~Ooguri and Y.~Ookouchi,
 ``Direct mediation of meta-stable supersymmetry breaking,''
 Phys.\ Rev.\  D {\bf 75}, 045022 (2007), 
 arXiv:hep-ph/0612139.
}

\lref\CST{
 C.~Csaki, Y.~Shirman and J.~Terning,
 ``A simple model of low-scale direct gauge mediation,''
 arXiv:hep-ph/0612241.
}

\lref\Nelson{
 A.~E.~Nelson and N.~Seiberg,
 ``R symmetry breaking versus supersymmetry breaking,''
 Nucl.\ Phys.\  B {\bf 416}, 46 (1994), 
 arXiv:hep-ph/9309299.
}

\lref\Dine{
 M.~Dine and J.~Mason,
 ``Gauge mediation in metastable vacua,''
 arXiv:hep-ph/0611312.
}

\lref\Shih{
 D.~Shih,
 ``Spontaneous R-symmetry breaking in O'Raifeartaigh models,''
 arXiv:hep-th/0703196.
}

\lref\AGM{
 A.~Amariti, L.~Girardello and A.~Mariotti,
 ``On meta-stable SQCD with adjoint matter and gauge mediation,''
 arXiv:hep-th/0701121.
}

\lref\vafa{
  C.~Vafa,
  ``Superstrings and topological strings at large $N$,''
  J.\ Math.\ Phys.\  {\bf 42}, 2798 (2001), 
  arXiv:hep-th/0008142.
}
\lref\CFIKV{
  F.~Cachazo, B.~Fiol, K.~A.~Intriligator, S.~Katz and C.~Vafa,
  ``A geometric unification of dualities,''
  Nucl.\ Phys.\  B {\bf 628}, 3 (2002), 
  arXiv:hep-th/0110028.
}

\newbox\tmpbox\setbox\tmpbox\hbox{\abstractfont }
\Title{\vbox{\baselineskip12pt \hbox{CALT-68-2642}\hbox{UT-07-12}}
}
{\vbox{\centerline{Gauge Mediation in String Theory}}}
\vskip 0.2cm

\centerline{Teruhiko Kawano,$^1$ Hirosi Ooguri,$^{1,2}$ 
and Yutaka Ookouchi$^2$}
\vskip 0.4cm
\centerline{$^1$\it
Department of Physics, University of Tokyo,
Tokyo 113-0033, Japan}
\vskip 0.2cm
\centerline{$^2$\it California Institute of Technology, Pasadena,
CA 91125, USA}

\vskip 1.3cm

\centerline{\bf Abstract}

\medskip

We show that a large class of phenomenologically viable  
models for gauge mediation of supersymmetry breaking 
based on meta-stable vacua can be realized 
in local Calabi-Yau compactifications of string theory.

\noindent

\bigskip\bigskip
\Date{April, 2007}

\vfill
\eject

\newsec{Introduction}

The use of meta-stable vacua in supersymmetric model building 
has attracted much attention lately, especially after the discovery 
 \ISS\ that generic supersymmetric field theories in four dimensions 
such as the supersymmetric QCD with massive flavors have 
meta-stable vacua with broken supersymmetry.
In \KOO , realistic models of direct mediation were
constructed using superpotentials without $U(1)_R$ 
symmetry. Though explicit breaking of the $U(1)_R$ symmetry 
generates meta-stable vacua, there is a range of parameters 
where one can make them sufficiently long lived, while satisfying
the phenomenological constraints on the masses of 
the gauginos, the gravitino, and the scalars without 
artificially elaborate constructions. 
The models can also avoid producing
Landau poles in standard model gauge interactions below the
unification scale.
Recently beautiful realizations of these models in string theory,
including a natural mechanism to generate small parameters of these
models, were found in \kachru . 
 
Gauge mediation models were also constructed using meta-stable vacua
with similar phenomenological benefits \refs{\Murayamaa,\Murayamab}. Related 
ideas have been explored in \refs{\AGM\Dine\CST\Aharony-\Shih}.
Accepting the possibility that our universe 
may be in a meta-stable state allows us to
circumvent the theoretical constraints due to the Nelson-Seiberg theorem 
on R-symmetry \ref\Nelson{
  A.~E.~Nelson and N.~Seiberg,
  ``R symmetry breaking versus supersymmetry breaking,''
  Nucl.\ Phys.\  B {\bf 416}, 46 (1994), 
  arXiv:hep-ph/9309299.
}
and the Witten index \ref\WittenNF{
  E.~Witten,
  ``Dynamical Breaking Of Supersymmetry,''
  Nucl.\ Phys.\  B {\bf 188}, 513 (1981).
} and gives us greater flexibility in model building, as emphasized
in \ref\IntriligatorPY{
  K.~Intriligator, N.~Seiberg and D.~Shih,
  ``Supersymmetry breaking, R-symmetry breaking and metastable vacua,''
  arXiv:hep-th/0703281.
} among other recent papers.

Among the models constructed recently based on meta-stable vacua, 
the ones discussed in \Murayamaa\ are particularly simple. 
In this paper, we will show that they have 
ultra-violet completions in supersymmetric quiver gauge theories 
which can be realized in string compactifications. 
Moreover, our construction can be naturally generalized to a large
class of quiver gauge theories, providing a basis for 
the speculation in \Murayamaa\ 
that ``gauge mediation may be a rather generic phenomenon in the landscape of 
possible supersymmetric theories.'' 
In this paper, we will demonstrate the idea by explicitly working out 
one example: a model based on type IIB superstrings compactified on 
the $A_4$-fibered geometry 
\ref\CKV{
  F.~Cachazo, S.~Katz and C.~Vafa,
  ``Geometric transitions and ${\cal N} = 1$ quiver theories,''
  arXiv:hep-th/0108120.
}. 
We will also give an outline of generalizations of this construction 
to a large class of quiver gauge theories. 
Detailed analysis of meta-stable vacua in these models
will be given in a separate paper 
\ref\KOOP{
  T. Kawano, H. Ooguri, Y. Ookouchi and C.S. Park,
   in preparation
}.

\newsec{The Model}

The model we will consider in this paper is realized in
string theory compactified on the local Calabi-Yau manifold
described by the equation, 
\eqn\Afour{ \eqalign{ & 
x^2 + y^2 + \prod_{i=1}^5 \left(z + t_i (w)
\right) = 0,  
\qquad
\sum_{i=1}^{5}t_i(w)=0, 
\cr
& t_i(w)-t_{i+1}(w) = \mu_i (w - x_i),  ~~~~ (x,y,z,w) \in \IC^4. 
}}
Since $t_i$'s are functions of $w$,
this gives the $A_4$ singularity fibered over $w \in \IC$. 
In particular, 
there exist four two-cycles $S^2$ on which D branes can be wrapped. 
The low energy limit of  D$5$ branes wrapping the two-cycles $S^2$ 
and extending along the four uncompactified dimensions is 
the $A_4$ quiver gauge theory
with the gauge group $U(N_1) \times U(N_2)\times U(N_3) \times U(N_4)$ 
with the adjoint chiral multiplets $X_{i=1,2,3,4}$ 
for the four gauge group factors and 
the bi-fundamental chiral multiplets $(Q_{12}, Q_{21})$, 
$(Q_{23}, Q_{32})$, and $(Q_{34}, Q_{43})$. 
This quiver gauge theory can also be realized on 
intersecting brane configuration with NS$5$ and D$4$ branes,
as expected from the T-duality between the $A_n$ singularity 
and NS$5$ branes 
\ref\OoguriWJ{
  H.~Ooguri and C.~Vafa,
  ``Two-dimensional black hole and singularities of CY manifolds,''
  Nucl.\ Phys.\  B {\bf 463}, 55 (1996), 
  arXiv:hep-th/9511164.
}.
 
\fig{\it $A_4$ quiver diagram
}{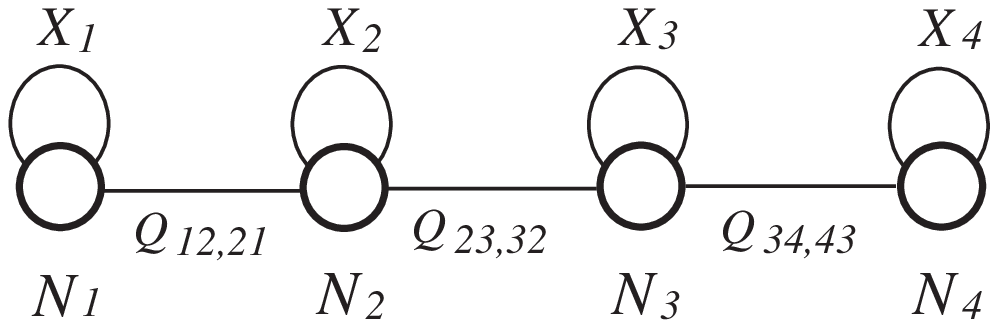}{5.5 truecm}
\figlabel\zuI

From the Calabi-Yau singularity \Afour, one can read off 
the superpotential of the quiver theory as \refs{\vafa,\CFIKV}
\eqn\WAfour{ 
W_{A_4}  = \sum_{i=1}^3 \tr \left( Q_{i+1,i} X_i Q_{i,i+1}
 - Q_{i,i+1} X_{i+1} Q_{i+1,i} \right) 
 + \sum_{i=1}^4 \tr\ {\mu_i \over 2} (X_i - x_i)^2. }
Note that the dimensionful parameters $\mu_i$ and $x_i$  
are the moduli of the Calabi-Yau manifold given by  \Afour  ,
namely they are closed string moduli.
The dynamical scales $\Lambda_{i=1,\cdots,4}$ of 
the four gauge group factors are also closed string moduli,
related to the sizes of the $S^2$'s. 
These closed string moduli are frozen and can be regarded as
parameters of the low energy theory. 
Let us suppose that $\mu_i$ are sufficiently larger than 
$\Lambda_i$ so that we can integrate out all the adjoints 
$X_i$ to obtain the effective superpotential 
\eqn\Weff{\eqalign{
W_{{\rm eff}}=&\sum_{i=1}^{3}m_i \tr\,Q_{i\,i+1}Q_{i+1\,i}
-\sum_{i=1}^{3}{1\over\tilde \mu_i} \tr\,\left(Q_{i\,i+1}Q_{i+1\,i}\right)^2
\cr
&~~~+{1\over\mu_2}\tr\,Q_{21}Q_{12}Q_{23}Q_{32}
+{1\over\mu_3}\tr\,Q_{32}Q_{23}Q_{34}Q_{43},
}}
where 
$$
m_i=c_i-c_{i+1}, 
\qquad 
\tilde \mu_i={2 \mu_i \mu_{i+1} \over \mu_i + \mu_{i+1}}
\qquad (i=1,2,3). 
$$

This quiver gauge theory can be used as a gauge mediation model as follows. 
We identify the bi-fundamentals $(Q_{34}, Q_{43})$ as messenger fields.
One way to incorporate the standard model sector would be  
to identify a subgroup of the $U(N_4)$ gauge group with 
the standard model gauge group or a GUT gauge group.
Alternatively, we can replace the 4th node of the quiver diagram 
of Fig. \zuI\ carrying the $U(N_4)$ gauge group 
with a string theory construction of the standard model.
For example, if the standard model is realized on intersecting
branes, messengers can be open strings connecting the 3rd node
carrying the $U(N_3)$ gauge group to the standard model branes.\foot{
In this case, we can still use the effective potential \Weff\ to describe
the interaction of the messengers and the hidden sector, but 
we should set ${1/\tilde\mu_4} = {1/(2\mu_3)}$ since
we do not have the adjoint field $X_4$.}     
In the following, we will denote the bi-fundamental fields $(Q_{34}, Q_{43})$
as  $(f,\tilde{f})$ to distinguish them from the rest of the quiver gauge
theory and to emphasize their role as the messengers.

The rest of the quiver gauge theory is treated as a hidden sector, 
where supersymmetry is broken dynamically. To use the result of \ISS,  
let us assume that the ranks of the gauge group factors satisfy 
\eqn\rankG{
N_2+1\leq N_1+N_3 < {3\over2}N_2
}
and that 
$$\Lambda_1, \Lambda_3, \Lambda_4\ll \Lambda_2 \ll \mu_i .$$
In this case, one can identify the gauge group $SU(N_c)$ of the model of \ISS\ 
with $SU(N_2)\subset U(N_2)$ of the quiver theory.
Since the metastable vacuum can be found near the origin of the meson fields 
$M_{11}\sim{}Q_{12}Q_{21}$, $M_{33}\sim{}Q_{32}Q_{23}$, the terms 
$\tr\left(Q_{12}Q_{21}\right)^2$, $\tr\left(Q_{32}Q_{23}\right)^2$ and 
$\tr\left(Q_{12}Q_{21}Q_{32}Q_{23}\right)$ in the superpotential \Weff\ 
are irrelevant in our discussion below, if the masses $\mu_i$ of the adjoints 
satisfy the following bounds \refs{\Murayamaa,\Murayamab},   
\eqn\supression{ 
{\Lambda_2^2 \over \tilde\mu_{1,2}},~{\Lambda_2^2 \over \mu_{2}}  
\quad\leq\quad \min\left\{
{1\over 4\pi}\sqrt{m_{1,2} \Lambda_2},\,{1\over
16 \pi^2}{m_3\mu_3\over\Lambda_2}\right\}.
}

In this range of the parameters, the hidden sector and 
its interaction with the messenger sector
is described by the superpotential,
\eqn\Wour{
W=m\tr\,Q_{12}Q_{21}+m\tr\,Q_{32}Q_{23}
+{1\over\mu_3}\tr\,Q_{32}Q_{23}f\tilde{f}
+m_3\tr\,f\tilde{f}-{1\over\tilde{\mu}_3}\tr\,\left(f\tilde{f}\right)^2.
}
Here, we set the mass parameters $m_1= m_2=m$, for simplicity. 
Consider the case when $N_1=N_2=3$ and $N_3=1$ so that the
Landau pole problem can easily be avoided.
The resulting model is a variant of 
the models proposed in \Murayamaa. 
The model \Murayamaa\ has the global 
symmetry $U(4)\times{U}(1)_{\rm mess}$, 
where $U(4)$ is the flavor symmetry of the ISS model
and ${U}(1)_{\rm mess}$ acts on the messengers
$(f,\tilde{f})$. The meta-stable vacuum spontaneously breaks 
the $U(4)$ symmetry, giving rise to Nambu-Goldstone bosons, 
when $m_1\simeq{m}_2$.
In our model, the would-be Nambu-Goldstone bosons are
eaten by the gauge symmetry. This difference
is not important in the low energy analysis of supersymmetry breaking
effects. 

Let us discuss phenomenological constraints on 
the parameters in \Wour .
We will focus on the following part of the superpotential \Wour ,
\eqn\Wmess{
W_{{\rm mess}}={\Lambda_2\over\mu_3}\,M_{33}f\tilde{f}+m_3\,f\tilde{f},
}
where $M_{33}=Q_{32}Q_{23}/\Lambda_2$
is neutral under the $U(N_3)=U(1)$ gauge group. 
We have dropped the irrelevant quartic term $(f\tilde{f})^2$ because 
the messengers $(f, \tilde{f})$ are weakly interacting at energies 
above the electroweak scale, if 
the mass parameter $\tilde\mu_3$ is large enough. 
The $F$-component of the 
meson superfield $M_{33}$ develops the vacuum expectation value and breaks 
supersymmetry \ISS. The supersymmetric mass and the soft supersymmetry 
breaking mass of the messenger fields $(f,\tilde{f})$ are then given by
\eqn\effWmess{\eqalign{
W_{{\rm mess}}\simeq\left(m_3+\theta^2{m\Lambda_2^2\over\mu_3}\right)f\tilde{f}. 
}}
Following the analysis in  \refs{\Murayamaa,\Murayamab}, we find that
all the phenomenological requirements for the messenger sector
can be satisfied, for example, in the following range of parameters,
\eqn\phenomconditions{
\eqalign{& \Lambda_2\simeq10^{11} {\rm GeV},~~~ m\simeq10^8 {\rm GeV}, ~~~
m_3\simeq10^7 {\rm GeV}, \cr
& \mu_1\geq\mu_2\geq10^{13} {\rm GeV}, ~~~ 
\mu_3\simeq10^{18} {\rm GeV}.}}


\newsec{Generalization}

We found that both the messenger sector and the hidden sector of the 
models proposed in \Murayamaa\ can be
realized in the $A_4$ quiver gauge theory. This construction naturally suggests
the following generalization. 
Consider a quiver diagram which can be separated into two disjoint diagrams
$\Gamma_1$ and $\Gamma_2$ by cutting at one node, which we denote by $a$.
If the scale $\Lambda_a$ associated to the gauge group on the 
$a$-node is sufficiently low, and if superpotential interactions between them 
are small, we have effectively two separate quiver 
gauge theories for phenomena much above the scale $\Lambda_a$, one associated 
to $\Gamma_1$ and another associated to $\Gamma_2$, which are weakly 
interacting with each other through the $a$-node.
If supersymmetry is broken in the sector $\Gamma_1$, it can 
be communicated to the sector $\Gamma_2$ by the gauge mediation mechanism.
The beauty of the quiver gauge theory construction is that, because of the
presence of bi-fundamental and adjoint fields on links and nodes, 
an effective superpotential of the form \effWmess\ is naturally generated
when supersymmetry is broken in a part of the diagram connected to the 
$a$-node. 

It follows trivially that any quiver theory that is vector like with
adjustable mass terms has meta-stable supersymmetry 
breaking vacua in some range of its parameter. 
All one has to do is to identify a part of the diagram where supersymmetry can 
be broken using a known mechanism, for example as in \ISS\ or its variant 
\ref\OO{  
H.~Ooguri and Y.~Ookouchi,
`Landscape of supersymmetry breaking vacua in geometrically realized gauge 
theories,'' Nucl.\ Phys.\  B {\bf 755}, 239 (2006), 
arXiv:hep-th/0606061.
},  and to have its effect communicated to the rest 
of the diagram by messengers. One can also consider the scenario where
the quiver theory associated to a sub-diagram $\Gamma_2$ has
a supersymmetric vacuum with dynamically generated small scales, 
which can be used to set parameters of the theory associated to 
another sub-diagram $\Gamma_1$, where supersymmetry is broken.
The supersymmetry breaking effect can then be communicated back to
the sub-diagram $\Gamma_2$. This would give a string theory realization
of the idea of \ref\DineGM{
  M.~Dine, J.~L.~Feng and E.~Silverstein,
  ``Retrofitting O'Raifeartaigh models with dynamical scales,''
  Phys.\ Rev.\  D {\bf 74}, 095012 (2006), 
  arXiv:hep-th/0608159.
}. These and other mechanisms of supersymmetry breaking
will be explored further in \KOOP .

These supersymmetry breaking quiver gauge theories can be coupled to 
the messenger sector. In fact,  as in the case of the $A_4$ model discussed
in the previous section, the messenger sector itself can be included
in quiver theories. If the messenger sector is attached at the end of
the quiver diagram,  the effective low energy superpotential always takes
the form \Weff .
Thus, one can see that the models in \Murayamaa\ and their generalizations
are robust and naturally appear in this large class of string compactifications.

\vskip 2cm

\centerline{\bf Acknowledgments}

\medskip

We thank D.~Berenstein, M.~Dine, R.~Kitano, J.~Marsano, C.~S.~Park, 
N.~Seiberg, M.~Shigemori, and T.~Watari for discussions.  
H.O. thanks the hospitality of the high energy theory group 
at the University of Tokyo at Hongo.

H.O. and Y.O. are supported in part by the DOE grant
DE-FG03-92-ER40701. The research of H.O. is also supported in part
by the NSF grant OISE-0403366 and by the 21st Century COE Program
at the University of Tokyo. Y.O. is also supported 
in part by the JSPS Fellowship for Research Abroad. The research of 
T.K. was supported in part by the Grants-in-Aid (\#16740133) and (\#16081206) 
from the Ministry of Education, Culture, Sports, Science, and Technology of 
Japan.

\listrefs
\end